\documentclass[runningheads]{llncs}
\usepackage{graphicx}
\usepackage{caption}
\usepackage{array}
\newcolumntype{C}[1]{>{\centering\let\\\arraybackslash\hspace{0pt}}m{#1}}
\usepackage{tikz}
\usetikzlibrary{arrows}
\usepackage{diagbox}
\usepackage{hyperref}
\usepackage{subcaption}
\usepackage{tabu}
\usepackage{listings}
\usepackage{amsfonts}
\usepackage{tabularx}
\usepackage{rotating}

\renewcommand\figurename{Figure}

\definecolor{javared}{rgb}{0.6,0,0} 
\definecolor{javagreen}{rgb}{0.25,0.5,0.35} 
\definecolor{javapurple}{rgb}{0.5,0,0.35} 
\definecolor{javadocblue}{rgb}{0.25,0.35,0.75} 
\definecolor{editorbg}{HTML}{F3F3F3} 

\lstdefinelanguage{SQL}
{
	morekeywords={
		AUTHOR,
		AUTHORS,
		PUBLICATIONS,
		PUBLICATION,
		STREAM,
		STREAMS,
		INSTITUTION,
		INSTITUTIONS,
		CITED\_BY,
		AUTHORS\_PUBLICATIONS
	},
	sensitive=false, 
	morecomment=[l]{//}, 
	morecomment=[s]{/*}{*/}, 
	morestring=[b]' 
}

\lstdefinelanguage{XML}
{
	morestring=[b]",
	morestring=[s]{>}{<},
	morecomment=[s]{<?}{?>},
	stringstyle=\color{black},
	identifierstyle=\color{javapurple},
	keywordstyle=\color{javapurple},
	morekeywords={xmlns,version,type}
}

\begin{document}
\title{SchenQL - A Domain-Specific Query Language on Bibliographic Metadata}
\subtitle{Extended Version}

%
\author{Christin Katharina Kreutz\orcidID{0000-0002-5075-7699} \and Michael Wolz\orcidID{0000-0002-9313-7131} \and
Ralf Schenkel\orcidID{0000-0001-5379-5191}}
\authorrunning{Kreutz et al.}
%
\institute{Trier University, 54286 Trier, DE\\
	\email{\{kreutzch, s4miwolz, schenkel\}@uni-trier.de}}
\maketitle              
\begin{abstract}
Information access needs to be uncomplicated, users rather use incorrect data which is easily received than correct information which is harder to obtain. Querying bibliographic metadata from digital libraries mainly supports simple textual queries. A user's demand for answering more sophisticated queries could be fulfilled by the usage of SQL. As such means are highly complex and challenging even for trained programmers, a domain-specific query language is needed to provide a straightforward way to access data.

In this paper we present SchenQL, a simple query language focused on bibliographic metadata in the area of computer science while using the vocabulary of domain-experts. By facilitating a plain syntax and fundamental aggregate functions, we propose an easy-to-learn domain-specific query language capable of search and exploration. It is suitable for domain-experts as well as casual users while still providing the possibility to answer complicated queries. A user study with computer scientists directly compared our query language to SQL and clearly demonstrated SchenQL's suitability and usefulness for given queries as well as users' acceptance. 

\end{abstract}
\section{Introduction}

Scientific writing almost always starts with thorough bibliographic research on relevant publications, authors, conferences, journals and institutions. While web search is excellent for query answering and intuitively performed, not all retrieved information is correct, unbiased and categorized \cite{yates}. The arising problem is people's tendency of rather using poor information sources which are easy to query than more reliable sources which might be harder to access \cite{bates}. This introduces the need for more formal and structured information sources such as digital libraries specialized on the underlying data which are also easy to query. Oftentimes, interfaces of digital libraries offer the possibility to execute search on all metadata or query attributes. In many cases, they are not suitable to fulfil users' information needs directly when confronted with advanced query conditions such as \textit{"Which are the five most cited publications written by $A$?"}. Popular examples of full-text search engines with metadata-enhanced search for a subset of their attributes \cite{beall} are dblp \cite{dblp} or semantic scholar \cite{semantic}. Complex relations of all bibliographic metadata can be precisely traversed using textual queries with restricted focus and manual apposition of constraints. While SQL is a standard way of querying databases, it is highly difficult to master \cite{xu}. Domain experts are familiar with the schema but are not experienced in using all-purpose query languages such as SQL \cite{hep,madaan}. Casual users of digital libraries are not versed in either.

To close this gap, we propose SchenQL, a query language (QL) specified on the domain of bibliographic metadata. It is constructed to be easily operated by domain-experts as well as casual users as it uses the vocabulary of digital libraries in its syntax. While domain-specific query languages (DSLs) provide a multitude of advantages \cite{borodin}, the most important aspect in the conception of SchenQL was that no programming skills or database schema knowledge is required to use it.

The contribution of this paper lies in the presentation of a domain-specific query language on bibliographic metadata in computer science which is the first to the best of our knowledge. It focuses on retrieval and exploration aspects as well as aggregate functions. The proposed QL is evaluated two-fold: 1) interviews with domain experts were used to find real applications as well as room for further development and 2) a quantitative user-study thoroughly evaluated effectiveness, efficiency and user satisfaction of the new DSL against SQL.

The remainder of this paper is structured as follows: In Section 2 related work is discussed before our dataset is presented in Section 3. Section 4 introduces the structure and syntax of SchenQL which is thoroughly evaluated in two parts in the following Section 5. The last Section 6 describes possible future research.

\section{Related Work}

Areas adjacent to the one we are tackling are \textit{search on digital libraries}, \textit{formalized query languages}, \textit{query languages deriving queries from natural language} and \textit{domain-specific query languages}.

With \textit{search on digital libraries}, several aspects need to be taken into consideration: The MARC format \cite{marc} is a standard for information exchange in digital libraries \cite{yates}. While it is useful for known-item search, topical search might be problematic as contents of the corresponding fields can only be interpreted by domain-experts \cite{yates}. Most interfaces on digital libraries provide field-based boolean search \cite{daffodil} which can lead to difficulties in formulating queries that require the definition and concatenation of multiple attributes. This might cause a substantial cognitive workload on the user \cite{berget}. Withholding or restriction of faceted search on these engines fails to answer complex search tasks \cite{beall}. We provide a search option on topical information which even casual users can operate while also offering the possibility to clearly define search terms for numerous attributes in one query. Faceted search is possible on almost all attributes.

Numerous works come from the area of \textit{formalized query languages}. Examples for SQL-like domain-unspecific QLs on heterogeneous networks are BiQL \cite{biql} which is suitable for network analysis and focuses on create/update functions or SnQL \cite{snql}, a social network query language specialized on transformation and construction operations. Other SQL-like QLs which are not restricted on a domain, focus on graph traversal while being document-centric \cite{sheng} or social network analysis while being person-centric \cite{socialite}. Some SQL-like query languages operate on RDF \cite{stil} or bibliographic databases using the MARC format \cite{lim}. While SchenQL is as structured as these formalized query languages, it does not depend on complicated SQL-like syntax. We argue that SQL is too complex to be effectively operated by casual users and hardly even by computer scientists. Our DSL is neither person-centric nor document-centric but both. Its focal point is the retrieval and exploration of data contrasting the creation/transformation focus of several of the presented languages.

Further research applies QLs in a way that users do not interact with it directly in using the system but in their back end. In many cases, graph-like structured data of heterogeneous networks is used to locate information semantically relevant to a unspecific query \cite{socialscope,unicorn}. Such an indirection could be a future step in the development of SchenQL.

Whether it be translation of natural language for the formulation of XQuery \cite{nalix}, the processing of identified language parts to build parse trees \cite{teli}, the representation of natural language queries posed to digital libraries in SPARQL \cite{bloehdorn} or the recent audio to SQL translation available for different databases and domains \cite{xu}, \textit{analysis of natural language and its conversion to machine processable queries} is an active field of research. Our proposed DSL does not translate natural language to SQL but offers a syntax which is similar to natural language.

\textit{Domain-specific query languages} come in many shapes. They can be SQL-like \cite{bio}, visual QLs \cite{hep,algovista} or use domain-specific vocabulary \cite{neuroql} but are typically specialized on a certain area. They also come in different complexities: For example MathQL \cite{mathql} is a query language in markup style on RDF repositories but a user needs to be mathematician to be able to operate it. The DSL proposed by \cite{madaan} stems from the medical domain and is designed to be used by inexperienced patients as well as medical staff. Naturally, there are hybrid forms: Some natural language to machine-readable query options are domain-specific \cite{rohil} and some DSLs might be transferable to other domains \cite{borodin}. With SchenQL, we want to provide a DSL which uses vocabulary from the domain of bibliographic metadata while being useful for experts as well as casual users.

\section{Dataset}

Contrary to attempts of modelling every particularity of bibliographic metadata as seen with MARC format \cite{marc} or Dublin Core, we concentrate on a few basic objects in our data model. In our concept, bibliographic metadata consists of persons and publications which they authored or edited. These persons can be affiliated with certain institutions. Manuscripts can be of type book, chapter, article, master or PhD thesis and are possibly published in different venues (conferences or journals). They can reference other papers and oftentimes are cited themselves. Figure \ref{citation} describes the difference between citations and references.

\definecolor{bl}{RGB}{0,112,192}
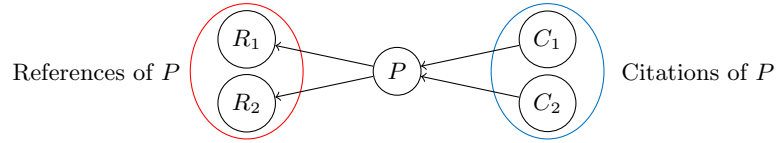
\begin{figure}[t]
	\centering
	\begin{tikzpicture}
	\node at (-2, 0.725)   (a) {References of $P$};
	\node at (6, 0.725)   (c) {Citations of $P$};
	\draw[red] (0,0.725) ellipse (0.75cm and 0.9cm);
	\draw[bl] (4,0.725) ellipse (0.75cm and 0.9cm);
	\node[draw, circle] at (0, 0.3)   (a1) {$R_2$};
	\node[draw, circle] at (0, 1.15)   (a2) {$R_1$};
	\node[draw, circle] at (2, 0.725)   (b) {$P$};
	\node[draw, circle] at (4, 0.3)   (c1) {$C_2$};
	\node[draw, circle] at (4, 1.15)   (c2) {$C_1 $};
	\draw [<-] (a1) -- (b);
	\draw [<-] (a2) -- (b);
	\draw [<-] (b) -- (c1);
	\draw [<-] (b) -- (c2);
	\end{tikzpicture}
	\caption{Nodes symbolize publications, edges between papers symbolize citations. $C_1$ and $C_2$ are citations of $P$, $R_1$ and $R_2$ are references of $P$.} 
	\label{citation}
\end{figure}

\begin{figure}[t]
	\includegraphics[width=\textwidth]{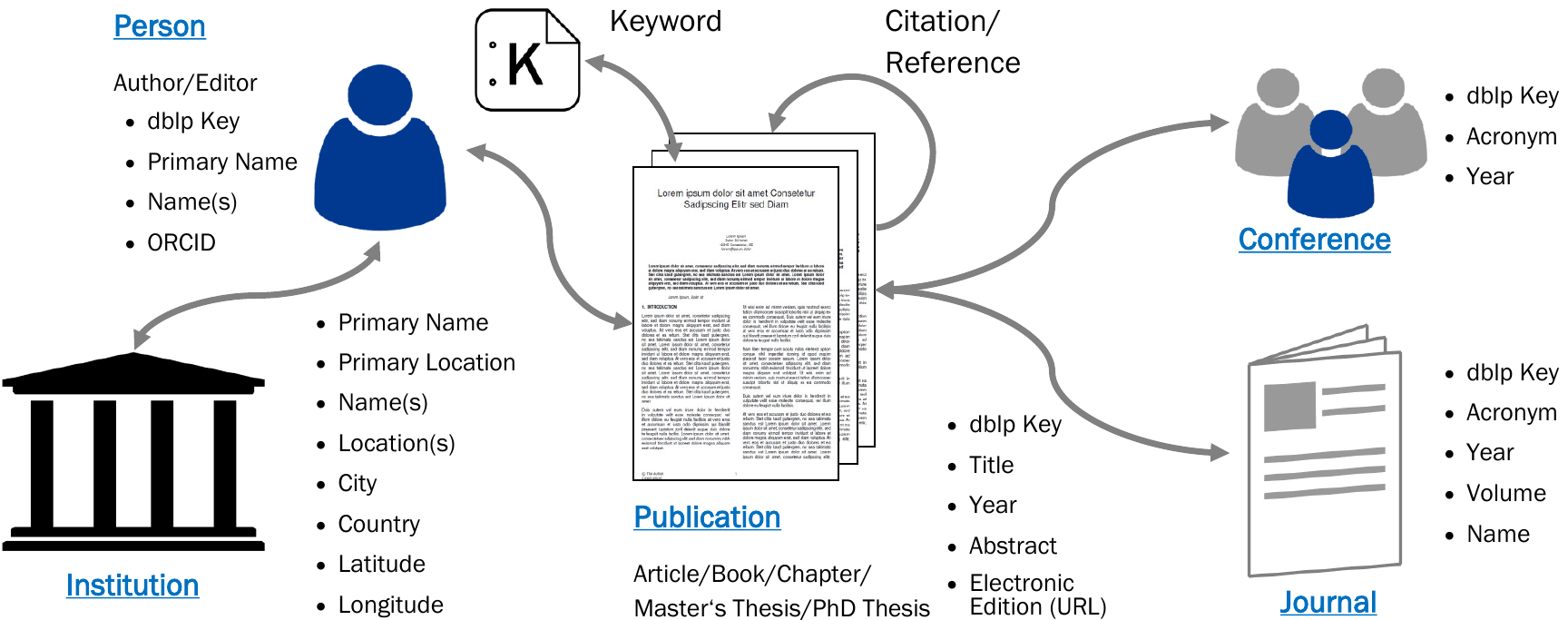}
	\caption{Relations, specializations and attributes of data objects from our extended dblp dataset.}
	\label{dataset}
\end{figure}

The dataset we evaluated on stems from the area of computer science. Our structures were filled with data from dblp \cite{dblp} mapped on data from SemanticScholar \cite{semantic} and enriched with information about institutions from Wikidata \cite{wikidata} with application specific extensions. As of June 2019, dblp contains data on more than 4.6 million publications, 2.3 million persons and several thousands of venues. Figure \ref{dataset} shows the relations, specializations and attributes of data objects in our dataset. 

\section{SchenQL}

The SchenQL Query Language was developed to pose the possibility to access bibliographic metadata in a textual manner which resembles natural language for casual as well as expert users of digital libraries in computer science. The fundamental idea in the development of the query language was to hide possibly complex operations behind plain domain-specific vocabulary. Such functionality would enable usage from anyone versed in the vocabulary of the domain without experience in sophisticated query languages such as SQL. SchenQL queries are formulated declarative, not procedural.

\subsection{Building Blocks}

\begin{table}[t]
\scriptsize
\setlength\tabcolsep{3pt}
\begin{tabularx}{\linewidth}{>{\hsize=0.025\hsize}X|>{\hsize=0.68\hsize}X|
			>{\hsize=0.51\hsize}X|
			>{\hsize=0.385\hsize}X|
			>{\hsize=0.51\hsize}X|
			>{\hsize=0.4\hsize}X} 
&\texttt{PUBLICATION} & \texttt{PERSON} & \texttt{CONFERENCE} & \texttt{JOURNAL} & \texttt{INSTITUTION}\\\hline 
\textbf{L} & dblp key, title & dblp key, primary name, orcid & dblp key, acronym & dblp key, acronym & \\\hline
\textbf{S} & \texttt{ARTICLE}, \texttt{MASTERTHESIS}, \texttt{CHAPTER}, \texttt{PHDTHESIS},  \texttt{BOOK}  
& {\texttt{AUTHOR}, \texttt{EDITOR}} &&&\\\hline
\textbf{F} & \texttt{PUBLISHED BY (I)}, \texttt{ABOUT (keywords)}, \texttt{WRITTEN BY (PE)}, \texttt{EDITED BY (PE)}, \texttt{APPEARED IN (C|J)}, \texttt{BEFORE year}, \texttt{IN YEAR year}, \texttt{AFTER year}, \texttt{TITLED title}, \texttt{REFERENCES (PU)}, \texttt{CITED BY (PU)}
&\texttt{PUBLISHED IN (C|J)}, \texttt{PUBLISHED WITH (I)}, \texttt{WORKS FOR (I)}, \texttt{NAMED name}, \texttt{ORCID orcid}, \texttt{AUTHORED (PU)}, \texttt{REFERENCES (PU)}, \texttt{CITED BY (PU)}
& \texttt{ACRONYM acronym}, \texttt{ABOUT (keywords)}, \texttt{BEFORE year}, \texttt{IN YEAR year}, \texttt{AFTER year}
& \texttt{NAMED name}, \texttt{ACRONYM acronym},\newline \texttt{ABOUT (keywords)}, \texttt{BEFORE year}, \texttt{IN YEAR year}, \texttt{AFTER year}, \texttt{VOLUME volume} 
& \texttt{NAMED name},\newline \texttt{CITY city}, \texttt{COUNTRY country}, \texttt{MEMBERS (PE)}\\\hline
\textbf{V} & title & primary name & acronym & acronym& primary name+primary location
\end{tabularx}
\vspace{5pt}
\caption{SchenQL base concepts \texttt{Publications} (\texttt{PU}), \texttt{persons} (\texttt{PE}), \texttt{conferences} (\texttt{C}), \texttt{journals} (\texttt{J}) and \texttt{institutions} (\texttt{I}) with their respective literals (L), specializations (S), filters (F) and standard return values (V).}
\label{schenql}
\end{table}

Base concepts are the basic return objects of the query language. A base concept is connected to an entity of the dataset and has multiple attributes. Those base concepts are \texttt{publications}, \texttt{persons}, \texttt{conferences}, \texttt{journals} and \texttt{institutions}. Upon these concepts, queries can be constructed. Base concepts can be specialized. For example \texttt{publications} can be refined by specializations \texttt{books}, \texttt{chapters}, \texttt{articles}, \texttt{master} or \texttt{PhD theses}. A specialization can be used instead of a base concept in a query.

Restrictions on base concepts are possible by using filters. A filter extracts a subset of the data of a base concept. Literals can be used as identifiers for objects from base concepts, they can be used to query for specific data. Attributes of base concepts can be queried. Table \ref{schenql} gives an overview of literals, specializations, filters and the standard return value for every base concept. In Figure \ref{dataset}, where base concepts are underlined and written in blue, their attributes are shown.

Functions are used to aggregate data or offer domain-specific operations. Right now, only three functions are implemented in SchenQL: \texttt{MOST CITED}, \texttt{COUNT} and \texttt{KEYWORDS OF}. The function \texttt{MOST CITED (PUBLICATION)} can be applied on publications. It counts and orders the number of citations of papers in the following set, and returns their titles as well as their number of citations. By default, the top 5 results are returned. \texttt{COUNT} returns the number of objects contained in the following subquery. \texttt{KEYWORDS OF (PUBLICATION | CONFERENCE | JOURNAL)} returns the keywords associated with the following base concept. The \texttt{LIMIT $x$} operator with $x \in \mathbb{N}$ can be appended at the end of any query to change the number of displayed results to $x$.
\subsection{Syntax}

The syntax of SchenQL follows some simple rules with the goal of being similar to queries formulated in natural language and therefore understandable and easy to construct. Queries are completed with a semicolon, subqueries have to be surrounded by parentheses. It is possible to write singular or plural when using base concepts or specializations (e.g. \texttt{CONFERENCE;} or \texttt{CONFERENCES;}). Filters follow base concepts or their specializations, can be in arbitrary order and are connected via conjunction. Most filters expect a base concept as parameter (e.g. \texttt{WRITTEN BY (PERSONS)}), several filters expect a string as parameter (e.g. \texttt{COUNTRY "de"}). Specializations can be used in place of base concepts. Instead of a query \texttt{PERSON NAMED "Ralf Schenkel";} a specialization like \texttt{AUTHOR NAMED "Ralf Schenkel";} would be possible.

If a filter requires a base concept, parentheses are needed except for the case of using literals for uniquely identifying objects of the base concept. For example \texttt{PUBLICATIONS WRITTEN BY "Ralf Schenkel";} is semantically equivalent to  \texttt{PUBLICATIONS WRITTEN BY (PERSONS NAMED "Ralf Schenkel");}.

\texttt{COUNT} can process any kind of subquery (e.g. \texttt{COUNT (INSTITUTIONS);}). \texttt{LIMIT $x$} can be appended to any query, \texttt{MOST CITED} requires a subquery which produces objects of base concept \texttt{PUBLICATION} (e.g. \texttt{MOST CITED (ARTICLES APPEARED IN "icadl") LIMIT 10;} returns the ten most cited articles which have appeared in the conference with acronym ICADL). \texttt{KEYWORDS OF} requires a subquery, which returns objects of type \texttt{PUBLICATION}, \texttt{JOURNAL} or \texttt{CONFERENCE}.
Attributes of base concepts can be accessed by putting the queried for attribute in quotation marks in front of a base concept and connecting both parts with an \texttt{OF} (e.g. \texttt{"dblpKey" OF JOURNALS IN YEAR 2010;})

\subsection{Implementation}

Our dataset is stored in a MySQL 8.0.16 database. Lexer and parser of the compiler were built using ANTLR \cite{antlr} with Java as target language. The compiler translates queries from SchenQL to SQL and runs them on the database. SchenQL can be used in a terminal client similar to the MySQL shell.

\section{Evaluation}

Our evaluation consists of two parts. In an initial, quantitative study we compared our domain-specific query language SchenQL against the all-purpose query language SQL. In the second, qualitative study, use-cases and possible further improvements were accessed. The quantitative study intended to measure the effectiveness and efficiency of SchenQL and users' satisfaction with it.

\subsection{Quantitative Study}

In the quantitative study, SQL was used in a terminal client as a widespread alternative query language to SchenQL, as it is not feasible to compare a specialized system to a commercial search engine and the differences between the compared systems should be minor \cite{eval}. The underlying data was stored in a MySQL database in version 8.0.16. Therefore, the requirement for test users was to be familiar with SQL.

\subsubsection{Queries}

\begin{table}[t]
	\centering
	\scriptsize
	\begin{tabular}{ll}
		$Q_1$ & What are the titles of publications written by author $A$?\\
		$Q_2$ & What are the names of authors which published on conference $C$?\\	
		$Q_3$ & What are the titles of the publications referenced by author $A$ in year $Y$?\\	
		$Q_4$ & What are the titles of the five most cited publications written by author $A$?\\
	\end{tabular}
\vspace{5pt}
	\caption{Overview of types of queries used in the evaluation, $A$ are authors, $C$ is a conference and $Y$ is a year.}
	\label{queries}
\end{table}

Our evaluation queries are inspired by everyday search tasks of users of digital libraries \cite{bloehdorn,pirolli}. We formulated four different types of queries targeting several base concepts and functionalities of SchenQL. Table \ref{queries} gives an overview of types of questions used in the evaluation. Variables were switched between query languages to prevent learning effects based on results of queries. $Q_1$, $Q_3$ and $Q_4$ are publication searches while $Q_2$ targets person search. 

A formulation of $Q_1$ in SQL would be:

\scriptsize
\begin{lstlisting}
SELECT title
FROM publication NATURAL JOIN person_authored_publication NATURAL JOIN person_names
WHERE person_names.name = "A";
\end{lstlisting}
\normalsize

In SchenQL, the same query could be expressed by the following:
\scriptsize
\begin{lstlisting}
PUBLICATIONS WRITTEN BY "A";
\end{lstlisting}
\normalsize

A formulation of $Q_2$ in SQL would be:

\scriptsize
\begin{lstlisting}
SELECT primaryName
FROM person NATURAL JOIN person_authored_publication NATURAL JOIN publication
WHERE publication.conferenceKey = "C";
\end{lstlisting}
\normalsize

In SchenQL, the same query could be expressed by the following:
\scriptsize
\begin{lstlisting}
AUTHORS PUBLISHED IN (CONFERENCE ACRONYM "C");
\end{lstlisting}
\normalsize

A formulation of $Q_3$ in SQL would be:

\scriptsize
\begin{lstlisting}
SELECT DISTINCT title
FROM publication p JOIN publication_references pr ON p.publicationKey = pr.pub2Key
WHERE pr.pub1Key IN ( 
  SELECT publicationKey
  FROM person_authored_publication NATURAL JOIN person_names NATURAL JOIN publication
  WHERE person_names.name = "A" AND year = Y
);
\end{lstlisting}
\normalsize
In SchenQL, the same query could be expressed by the following:
\scriptsize
\begin{lstlisting}
PUBLICATIONS CITED BY (PUBLICATIONS WRITTEN BY "A" IN YEAR Y);
\end{lstlisting}
\normalsize

And lastly, a formulation of $Q_4$ in SQL would be:

\scriptsize
\begin{lstlisting}
SELECT title, COUNT(*)
FROM publication p NATURAL JOIN person_authored_publication NATURAL JOIN person_names JOIN publication_references pr ON p.publicationKey = pr.pub2Key
WHERE person_names.name = "A"
GROUP BY title
ORDER BY COUNT(*) DESC
LIMIT 5;
\end{lstlisting}
\normalsize

In SchenQL, the same query could be expressed by the following:
\scriptsize
\begin{lstlisting}
MOST CITED (PUBLICATIONS WRITTEN BY "A");
\end{lstlisting}
\normalsize

As SQL queries tend to become complex relatively fast, the construction of more sophisticated queries was omitted.

\subsubsection{Setting}

\begin{sidewaysfigure}	
	\includegraphics[width=\textwidth]{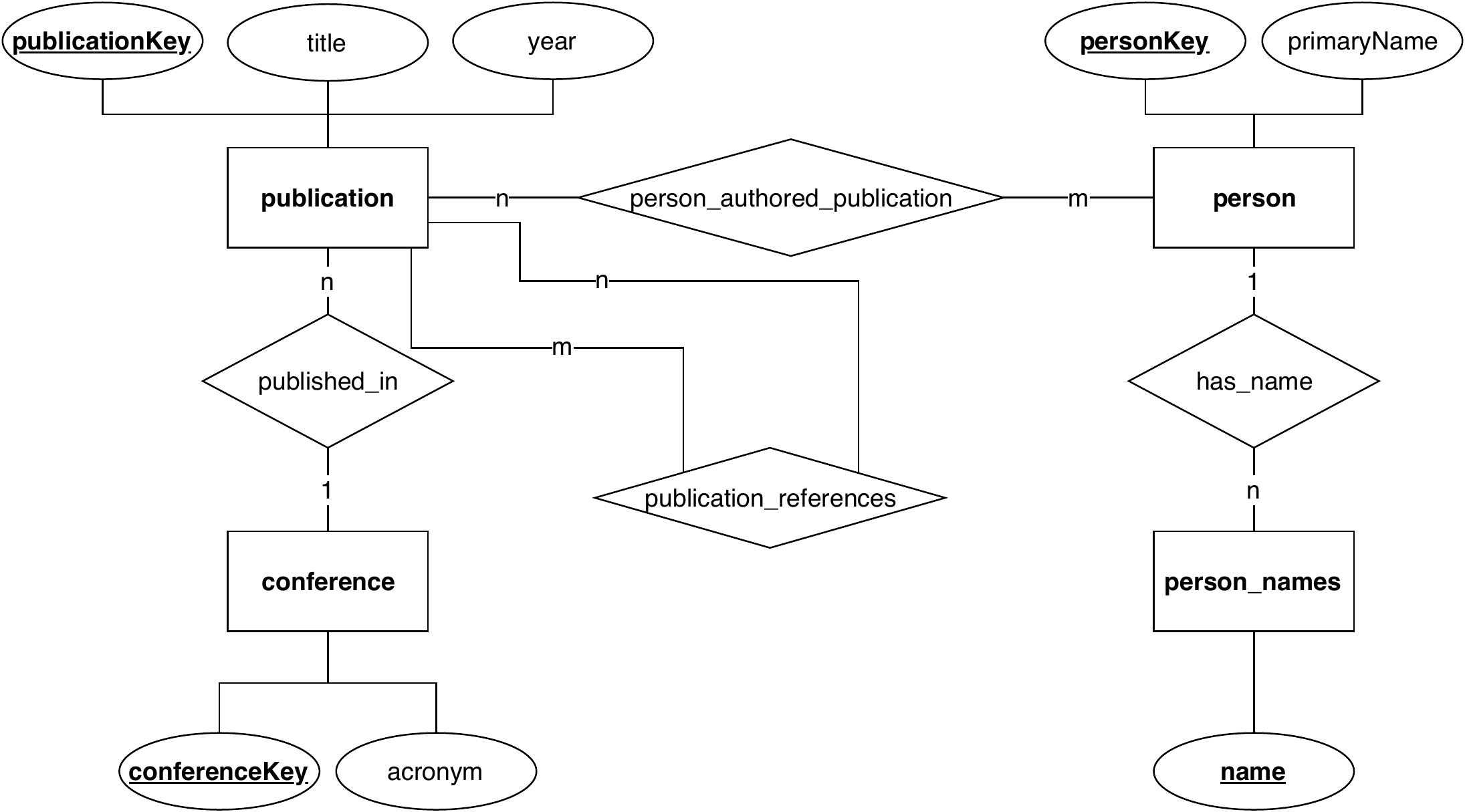}
	
	\caption{ER diagram of our used dataset.}
	\label{diagram}
\end{sidewaysfigure}

\begin{sidewaysfigure}	
	\includegraphics[width=\textwidth]{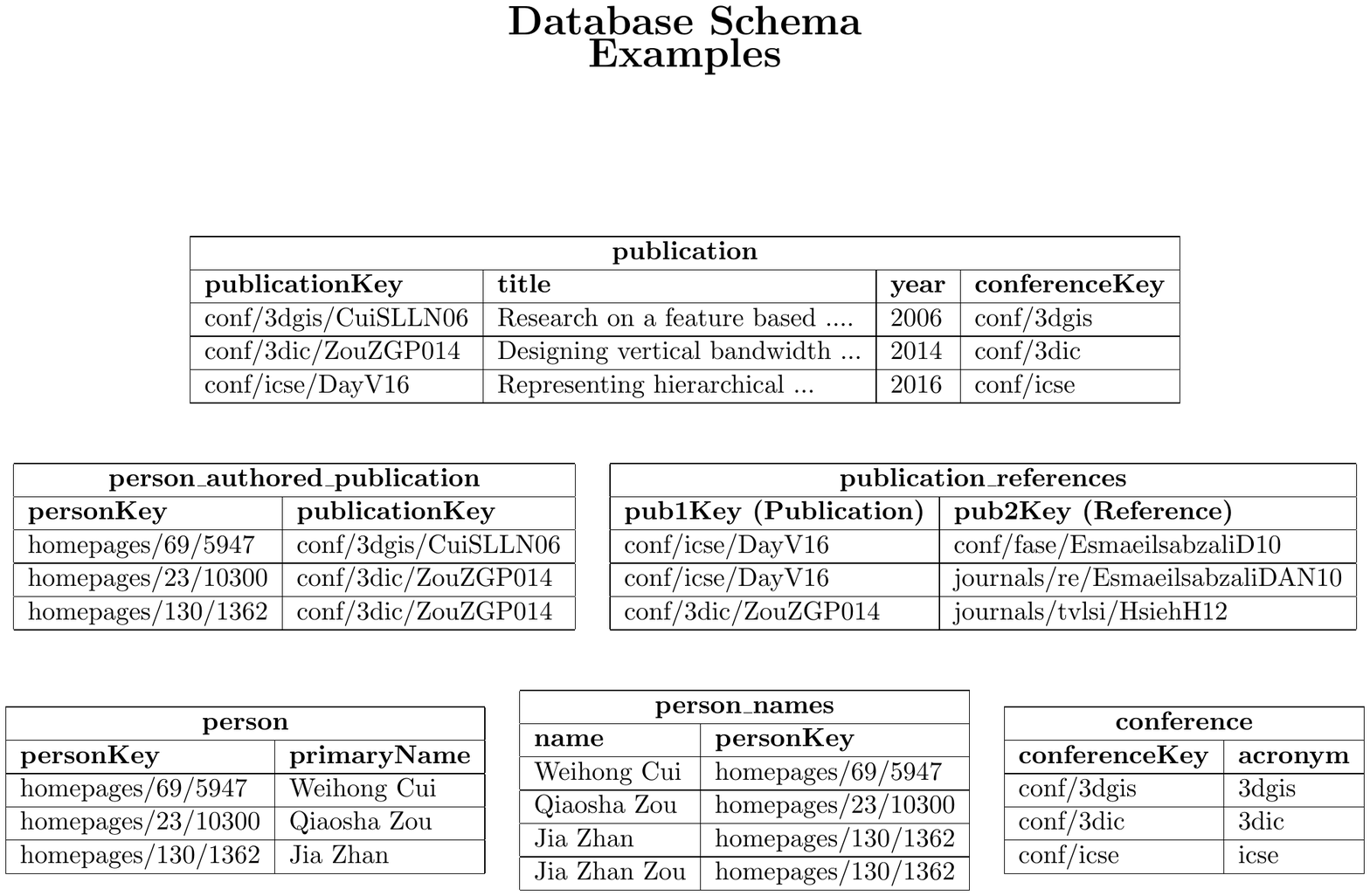}
	
	\caption{Schema of the database filled with example data.}
	\label{schema}
\end{sidewaysfigure}

\begin{sidewaysfigure}	
	\includegraphics[width=\textwidth]{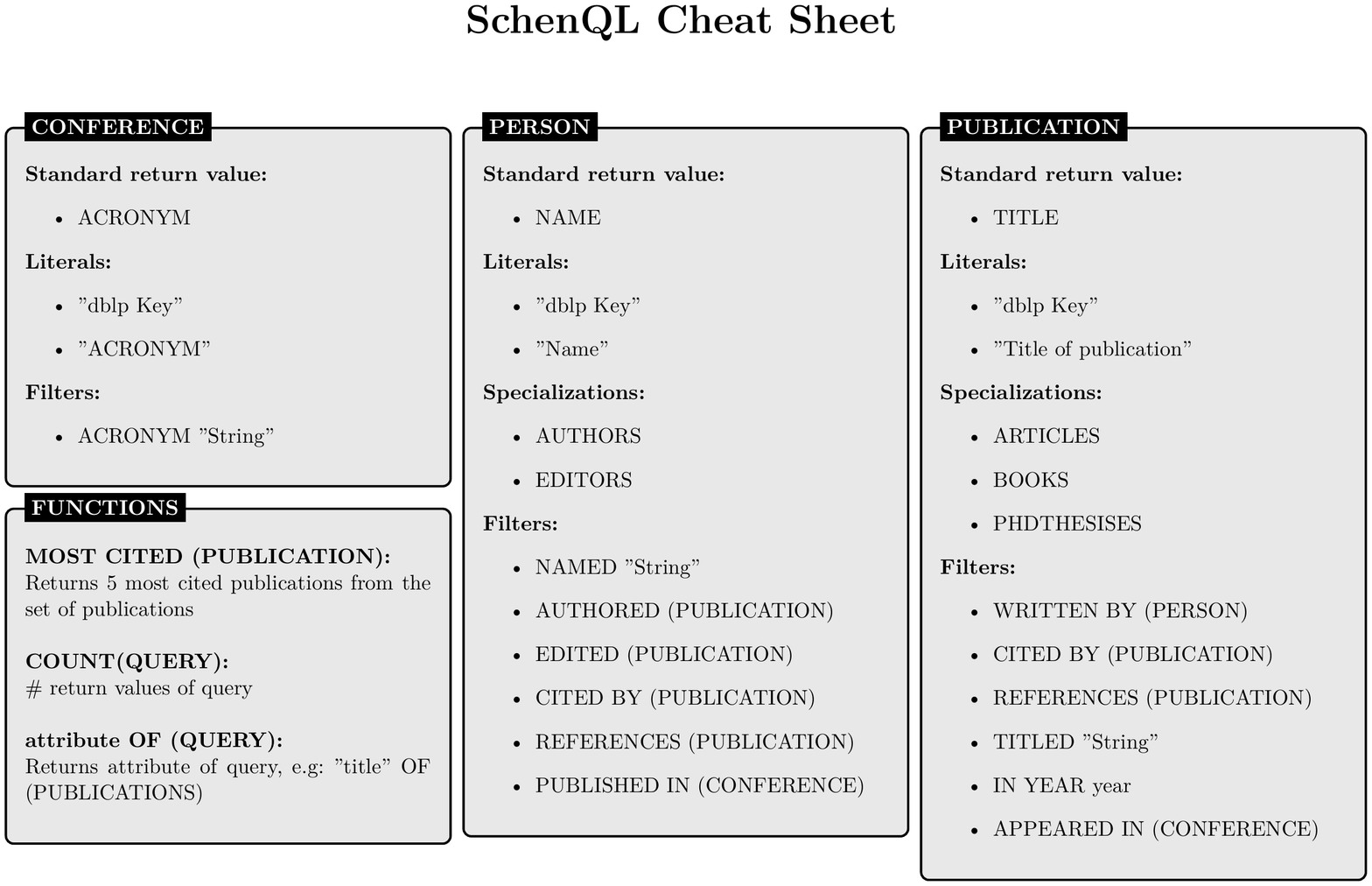}
	
	\caption{SchenQL cheat sheet.}
	\label{cheat}
\end{sidewaysfigure}

After a preliminary study with two participants, we defined the evaluation process of our archetypical interactive information retrieval study \cite{eval} as follows: Every user performed the evaluation alone in presence of a passive investigator on a computer with two monitors. The screens were captured in order to measure times used to formulate the queries. A query language, with which a user was going to begin the evaluation was assigned. Users were permitted to use the internet at any stage of the evaluation. For the evaluation, the presented tables in the database were restricted to the ones which were needed for formulation of the queries as the full ER diagram of the database could overwhelm users. A similar strategy was executed with SchenQL. Not all base concept, attributes and specializations were explained to test subjects but only a smaller subset which was roughly equivalent to the specified tables in SQL. Test users were given the ER diagram which is shown in Figure \ref{diagram}, Figure \ref{schema} which shows examples of data in the database schema and a SchenQL cheat sheet as shown in Figure \ref{cheat}. 

At first, a video tutorial \cite{video} for the introduction and usage of SQL and SchenQL was shown, afterwards subjects were permitted to formulate queries using the system they were starting to work with. Following this optional step, users were asked to answer a first online questionnaire to assess their current and highest level of SQL knowledge, the number of times they used SQL in the last three months as well as their familiarity with the domain of bibliographic metadata. The next part was the formulation of the four queries and their subjective rating of difficulty in the first query language before the query language was switched. Test users starting with SQL are contained in group A, those beginning with SchenQL are part of group B. Participants were asked to submit the queries in SQL and SchenQL respectively.

The evaluation was concluded with a second online questionnaire regarding the overall impression of SchenQL, the rating of SchenQL and SQL for the formulation of queries as well as several open questions targeting possible advantages and improvements of SchenQL.

We evaluated 21 participants (23 counting the subjects of the preliminary study) from the area of computer science. As the system to start with was rotated between users, ten subjects started by using SQL while eleven participants began the evaluation using SchenQL. We assume gender does not influence the measured values so it is not seen as additional factor in the evaluation \cite{eval}.

\subsubsection{Numerical Results}

\begin{table}[t]
	\centering
	\begin{tabu}{l|[1.5pt]l|l|[1.5pt]l|l|[1.5pt]l}
		& \multicolumn{2}{c|[1.5pt]}{\textbf{SQL}} & \multicolumn{2}{c|[1.5pt]}{\textbf{SchenQL}} & Difference \\
		Query & CORR & DIFF & CORR & DIFF & in DIFF\\\tabucline[2pt]{-}
		$Q_1$ & 90.48 & 2.85 ($\sigma$ = 1.77) & 90.48 & 1.57 ($\sigma$ = 0.68)&1.28\\
		$Q_2$ & 90.48 & 3 ($\sigma$ = 1.61) & 100 & 2.1 ($\sigma$ = 1.38)&0.9\\
		$Q_3$ & 23.81 & 4.86 ($\sigma$ = 1.31) & 47.62 & 2.71 ($\sigma$ = 1.19)&2.15\\
		$Q_4$ & 23.81 & 5.9 ($\sigma$ = 1.14) & 95.24 & 1.71 ($\sigma$ = 0.72)&4.19            
	\end{tabu}
\vspace{5pt}
	\caption{Correctness (CORR) in percent, assessed average difficulty (DIFF) and difference of average difficulty of the four queries for SQL and SchenQL. }
	\label{correctness}
\end{table}

Participants needed about an hour to complete the whole evaluation. All values are rounded on 2 decimal places. 

57.14\% of queries were correctly formulated using SQL whereas 83.33\% of queries were correctly formulated while using SchenQL. This result clearly shows the superior effectiveness of SchenQL compared to SQL. Table \ref{correctness} gives an overview of correcteness and average rated difficulty of all four queries for both languages. Difficulty was rated on a scale from 1 (very easy) to 7 (very difficult). While $Q_1$ and $Q_2$ were answered correctly by most participants, the number of correctly formulated queries for $Q_3$ and $Q_4$ highly depends on the system. While $Q_4$ was correctly answered by a quarter of subjects using SQL, more than 95\% of users were able to formulate the query in SchenQL. The mean rating of the formulation of queries with SQL was 4.15 ($\sigma$ = 1.94), with SchenQL the average rating was 2.02 ($\sigma$ = 1.11). On average, query construction using SQL is rated more difficult for every query. The highest rated difficulty for a query in SchenQL is still lower than the lowest rated difficulty of a query in SQL. 

\begin{figure}[t]
	\begin{minipage}{0.49\textwidth}
		\includegraphics[width=1\textwidth]{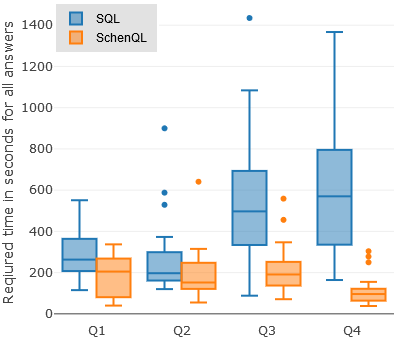}
	\end{minipage}
	\hfill
	\begin{minipage}{0.49\textwidth}
		\includegraphics[width=1\textwidth]{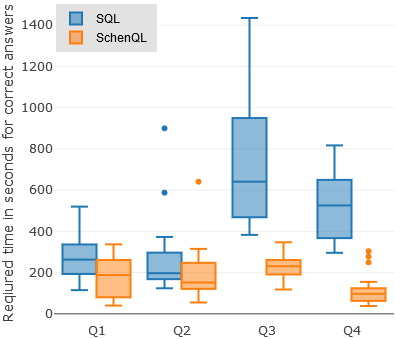}
	\end{minipage}
	\caption{Required time for all (on the left) and only correct (on the right) answers for all queries.}	
	\label{time}
\end{figure}

Average construction of queries in SQL took 7:15 minutes ($\sigma$ = 4:47 min.), in SchenQL the construction took 2:52 minutes ($\sigma$ = 1:51 min.) on average which documents the efficiency of our proposed DSL. Figure \ref{time} gives an overview of required times of participants for each query and query language. The significant difference of required time for $Q_3$ and $Q_4$ with SQL compared to the first two queries and the query construction times in SchenQL is remarkable. Both of these queries are assumed to be complex which is supported by the low percentage of correctly formulated queries using SQL. They are also much longer than the SchenQL queries so the time required to write them down is higher and there is more opportunity to make mistakes which causes query reformulation \cite{daffodil}.

\begin{table}[t]
	\centering
	\begin{tabu}{l|l|l|l}
		Correlation& CQ all & CQ SQL & CQ SchenQL\\\hline
		Current SQL skill&0.35&0.2&0.28\\
		Highest SQL skill&0.11&0.14&0\\
		Frequency of SQL&0.65&0.57&0.29\\		
	\end{tabu}
\vspace{5pt}
	\caption{Pearsons correlation coefficients for different combinations of measures of skill in SQL and number of correctly formulated queries (CQ) for every user.}
	\label{corr}
\end{table}

As the Pearson correlation coefficient between the fact that subjects were domain-experts and their frequency of usage of SQL in the last three months was 0.44, we assume that users from the domain of bibliographic metadata are also more experienced in using SQL. Table \ref{corr} shows correlations between skills of SQL, the frequency in which SQL was used in the last three months and the number of correctly formulated queries. The observed positive correlations lead to the assumption that experience in SQL helps in constructing queries with SchenQL which is compliant with the statement of users' previous experience with query formulation and bibliographic metadata impacting interactions with evaluated systems \cite{eval}.

For the following hypotheses tests, Welch' t-tests \cite{welch} with $p = 0.1$ were conducted. The null hypotheses of equal averages of required times to formulate queries as well as equal averages of ratings of difficulties of queries for the different QLs was rejected for each query $Q_1$, $\dots$, $Q_4$.

Of hypotheses concerning the average equality of required time to formulate queries, rated difficulty and correctness of queries of group A and group B, significant differences were only found in the equality of averages of required time for the formulation of $Q_1$ in SchenQL. Participants in group B were allowed to test queries in SchenQL before the evaluation started so this difference in data is not surprising. We were not able to observe between-subject learning effects.

Of hypotheses concerning the average equality of times, ratings of difficulty of queries and correct answers of the group of persons with prior experience in bibliographic metadata and those without previous knowledge, hypotheses concerning time needed for query formulation for $Q_3$ and $Q_4$ in SQL as well as the rating of difficulty of $Q_4$ in SchenQL were rejected. The average time for the twelve subjects experienced with bibliographic metadata for $Q_3$ was 11:02 minutes ($\sigma$ = 5:41 min., 3 correct answers), while the nine inexperienced users only took about 6:04 minutes ($\sigma$ = 2:51 min., 2 correct answers) to formulate the query. With $Q_4$ we observed a contrasting behaviour, domain-experts needed 8:28 minutes ($\sigma$ = 4:27 min., 4 correct answers) on average to complete the query, non-experts were much slower (13:26 min., $\sigma$ = 6:17 min., 1 correct answer). We suspect that domain-experts tend to review the result of their query online and therefore need more time to answer $Q_3$ than non-experts. Another explanation could be that since they are experienced with the principle of citations they were more confused with the one needed table as it contains publications and their references instead of two tables for papers, one which holds its citations and one which holds its references. For $Q_4$, the positive correlation between experience with bibliographic metadata and skill in SQL of a subject might be an explanation for the faster formulation of the complex query.

\begin{table}[t]
	\centering
	\begin{tabu}{l|l|l|l}
		User group&CORR in general&CORR SQL&CORR SchenQL\\\hline
		Non-experts&63.89& 47.22&80.56	\\
		Domain-experts&75 & 64.58 & 85.41 \\
	\end{tabu}
\vspace{5pt}
	\caption{Comparison of correctness (CORR) in percent for all queries and queries seperated by system for domain-experts as well as non-experts.}
	\label{be}
\end{table}

While observing the ratings of difficulty of $Q_4$ in SchenQL, the group of users without prior knowledge in the domain gave an 1.33 ($\sigma$ = 0.5). Experienced users rated the difficulty as 2 ($\sigma$ = 0.74) on average. This observation might stem from experienced subjects having certain expectations on how to formulate a query finding the most cited publications. They probably did not formulate this query for the first time but their expectations are not completely compliant to our query option. In general, 75\% of queries were correctly formulated from domain-experts whereas non-experts achieved only 63.89\%. Table \ref{be} shows the different percentages of correctness separated by system and user group.

The average overall impression of SchenQL was rated by the subjects as 5.05 ($\sigma$ = 0.74) on a scale from 1 (very bad) to 6 (very good). The rating of SchenQL for the formulation of the queries resulted in an average of 6.43 ($\sigma$ = 0.6) while the rating was 3.14 ($\sigma$ = 1.2) for SQL on a scale from 1 (very bad) to 7 (very good). While SQL was rated below mediocre, SchenQL was evaluated as excellent which shows the users high satisfaction with the proposed DSL.

\subsubsection{Open Questions}

In the concluding open questions, the short, easy and intuitive queries of SchenQL were complimented by many. Multiple users noted the comprehensible syntax was suitable for beginners as it is similar to natural language. Especially the usability for non computer scientists was emphasized.

Subjects criticized the need for a semicolon at the end of a query as unintuitive and wished for its abandonment as well as the parentheses for nested queries. Individuals noted their initial confusion due to the syntax and their incomprehension of usage of literals or limitations. Others asked for autocompletion, syntax highlighting, a documentation and more functions such as most cited with variable return values.

\subsection{Qualitative Study}

As the target audience of SchenQL consists of users of digital libraries, we performed interviews with four domain-experts from the dblp team to discover real-life requirements and use-cases as well as desirable extensions for our language or a possible future interface build on top of it.

Queries the domain-experts wanted to formulate in SchenQL included the computation of keywords of other publications which were published in the same journal as a given publication, the determination of the most productive or cited authors as well as the most cited authors with few co-authors. Except for the first query, users would not be able to construct these queries with SchenQL as most aggregate functions were out of scope for this paper. In SchenQL, the first query could be formulated as:

\scriptsize
\begin{lstlisting}
KEYWORDS OF (JOURNAL OF (PUBLICATIONS TITLED "P"));
\end{lstlisting}
\normalsize

According to domain-experts, an interface build on top of SchenQL should contain numerous visualizations: colour coded topics of publications or co-author-groups were wished for. Another subject requested building blocks for the visualization of graphs to display co-publications, co-institutions or neighbourhood of venues. Other desired functionalities of an interface would be a fault-tolerant person name search and sophisticated ranking methods.

\subsection{Discussion}

In terms of usability, the quantitative user study clearly showed the effectiveness and efficiency of SchenQL. More queries were answered correctly and much faster by using the DSL compared to SQL. The rating of difficulty of every query and the rating of overall impression suggest the users' satisfaction using SchenQL. Our user study showed the need for and adoption of the unknown query language and its usefulness for experts as well as non-experts in the specified domain of bibliographic metadata in computer science.

As we were unable to observe significant differences in the number of correct answers per query for domain-experts and non-experts, we assume SchenQL is suitable for both user groups. This assumption is additionally undermined by inability to distinguish between required times to formulate queries between the two groups when using the DSL. Tian et al. \cite{neuroql} stated that for a domain-expert, it would be easier to write queries in a DSL than in SQL, we measured this phenomenon with all test subjects.

Regarding learning effects related to the domain or fatigue, we did not observe significant differences in group A and group B's number of correct answers for the different queries which would indicate such cohesion.

The slight problems users had with the syntax of SchenQL could be bypassed by a GUI which supports the query process.

The qualitative study showed the domain-experts' need for more functions and a graphical user interface. While most of the proposed functions can be implemented in SchenQL, a GUI should have the option to formulate textual queries as well as visualize and manipulate the returned data.

\section{Conclusion and Future Work}

We proposed SchenQL, a domain-specific query language operating on bibliographic metadata from the area of computer science. Our thorough evaluation against SQL showed the need for such a DSL as well as the test subjects' satisfaction with it (overall average rating 5.05 on scale from 1 (very bad) to 6 (very good) and average rating for query formulation 6.43 on scale from 1 (very bad) to 7 (very good) for SchenQL). SchenQLs effectiveness and efficiency clearly surpasses SQL when applied in our domain, while 83.33\% of test queries were correctly formulated using SchenQL, only 57.14\% of queries were properly constructed with SQL. The average time needed to formulate a query with SchenQL was 4:23 minutes shorter than the mean time needed when using SQL. Our qualitative evaluation introduced the wish for more functions in SchenQL and an additional graphical user interface with numerous visualizations.

Possible improvements of SchenQL include adding logical set operations and incorporating of more data such as CORE rank \cite{core}, keywords derived from doc2vec \cite{d2v} or bibliometric measures such as h-index \cite{hindex}. More advanced enhancements of  functionalities of SchenQL could include the calculation of the influence graph of publications, centrality of authors, the length of a shortest path between two authors and the introduction of aliases for finding co-authors or co-citations as proposed in \cite{biql}. Algorithms for social network analysis as PageRank, computation of mutual neighbours, hubs and authorities or connected components could be worthwhile \cite{socialite}. As user-defined functions \cite{hep,socialite} and support in query formulation were well-received in other works \cite{daffodil}, they are a further prospect. The incorporation of social relevance in the search and result representation process as shown in \cite{socialscope,unicorn} would also be a possibility for extension. Via user profiles, written papers and interesting keywords could be stored, which in terms influence results of search and exploration.

SchenQL could also be used as an intermediate step between a user interface and the underlying database \cite{socialscope,unicorn}. Operability for casual users would be preserved by hiding future, more complex queries in visualization options in addition to a SchenQL query field.

%
%
%
%

\end{document}